\documentclass[article,twocolumn,amsmath,floats,showpacs,nofootinbib]{revtex4}
\usepackage{graphicx}% Include figure files
\usepackage{textcomp}

\usepackage{hyperref}

\hypersetup{colorlinks=true}

\begin{document}

\title{Point contact spectroscopy of the high-temperature superconductor  $La_{1.8}Sr_{0.2}CuO_{4}$}
\author{I.~K.~Yanson, L.~F.~Rybal'chenko, V.~V.~Fisun, N.~L.~Bobrov, M.~A.~Obolenskii*,
N.~B.~Brandt**, V.~V.~Moshchalkov**, Yu.~D.~Tret'yakov**, A.~R.~Kaul'**, and I.~E. ~Graboi**}
\affiliation{Physicotechnical Institute of Low Temperatures, Academy of Sciences of the Ukrainian SSR, Kharkov,\\
A. M. Gorky State University, Kharkov*,\\
and M. V. Lomonosov State University, Moscow**\\
Email address: bobrov@ilt.kharkov.ua}
\published {(\href{http://fntr.ilt.kharkov.ua/fnt/pdf/13/13-5/f13-0557r.pdf}{Fiz. Nizk. Temp.}, \textbf{13}, 557 (1987)); (Sov. J. Low Temp. Phys., \textbf{13}, 315 (1987)}
\date{\today}

\begin{abstract}The energy gap has been measured and the characteristic phonon energies determined for the superconducting compound $La_{1.8}Sr_{02}CuO_4$ from the current-voltage characteristics of point contacts with a copper counterelectrode. The temperature dependence of the energy gap was determined.
\end{abstract}
\maketitle

New superconducting compounds have recently been synthesized \cite{1,2} with critical temperatures considerably greater than the limit reached for compounds with the A15 structure. Specimens of the new material are baked ceramics of inhomogeneous composition: the superconductivity in them has a percolation character, which produces an appreciable smearing out of the superconducting transition. It is well known that the energy gap and the characteristic phonon frequencies of a superconductor can be determined from an investigation of the current-voltage characteristic (IVC) of point contacts of the S-c-N type. The information obtained in this way refers to a region of the material with dimensions of the order of the contact diameter $d$, immediately adjacent to the point contact.
Point contact spectroscopy is thus well suited to the study of heterophase systems: it enables the properties of separate superconducting clusters to be studied which have dimensions no greater than a few hundred \AA ngstroms.
\begin{figure}[]
\includegraphics[width=8cm,angle=0]{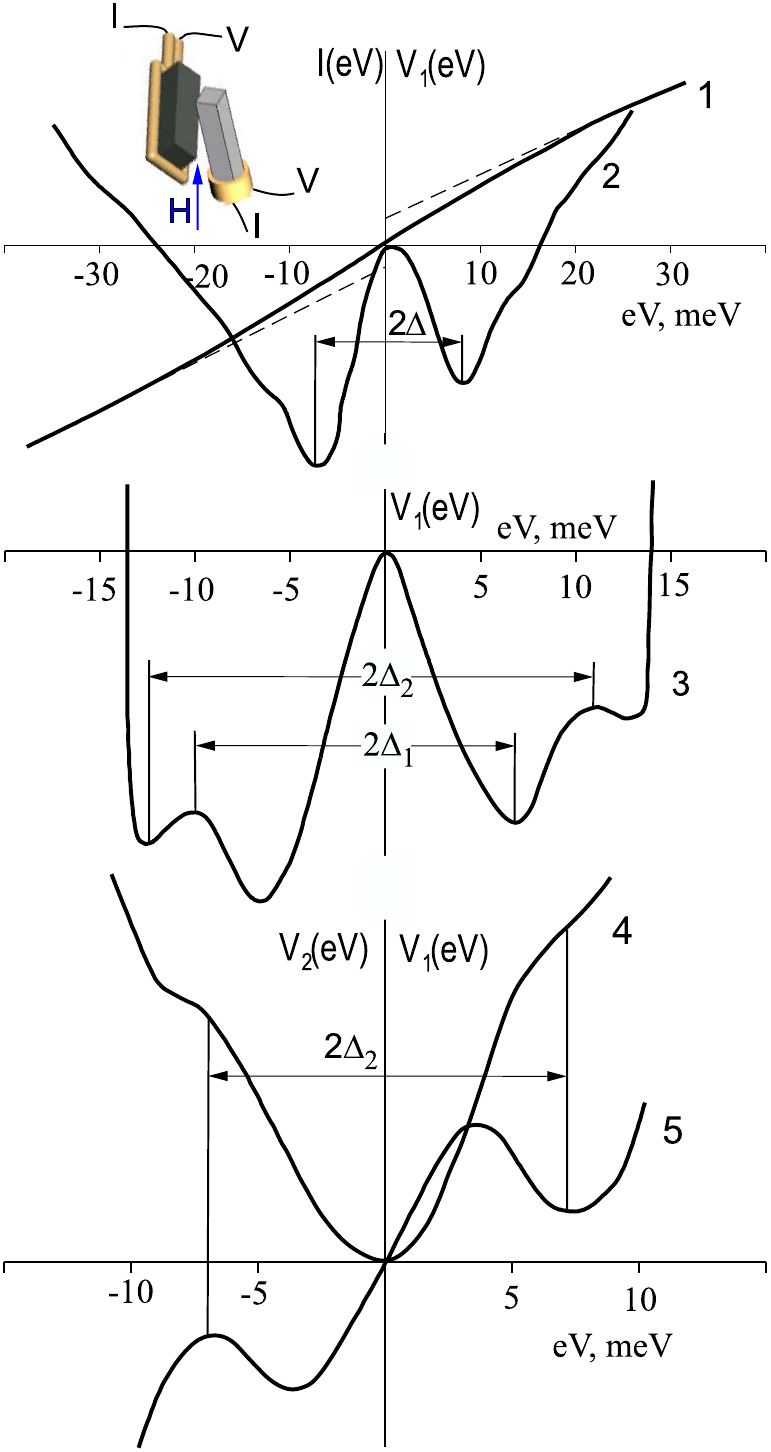}
\caption[]{Current-voltage characteristic and its derivatives for the hetero-junction $C2-Cu (C2\equiv La_{1.8}Sr_{0.2}CuO_{4})$: 1, 2)-$I(V)$, $V_1\sim dV/dI$, $T=9.3~K$,
$R_N=95~\Omega$, $2\Delta=15~meV$; 3-5) $R_N=750~\Omega$; 3) $T=1.5~K$, $2\Delta_1=13.3~meV$, $2\Delta_2=26~meV$; 4, 5) $T=14.8~K$, $2\Delta=15.5~meV$.}
\label{Fig1}
\end{figure}

\begin{figure}[]
\includegraphics[width=8cm,angle=0]{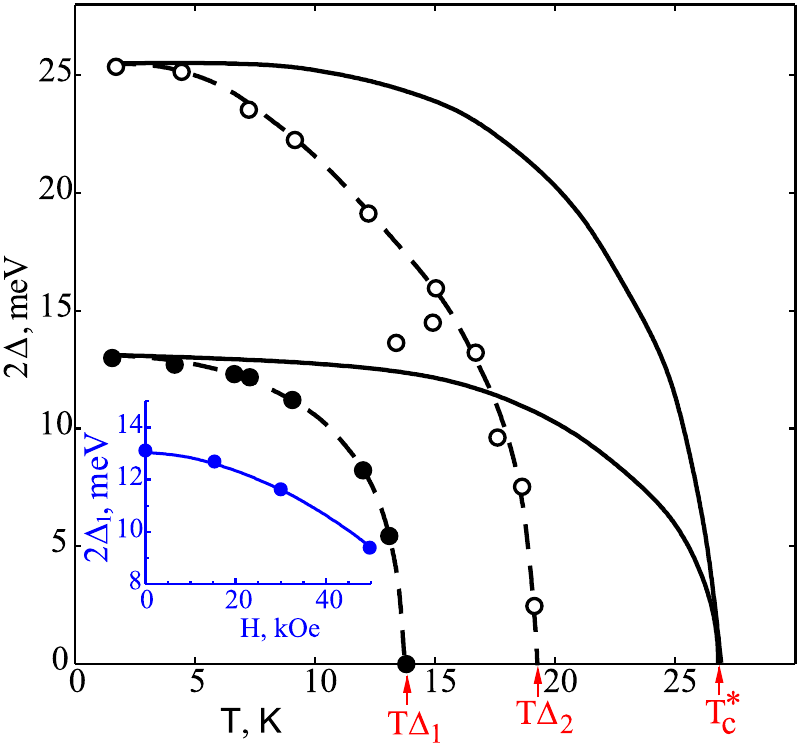}
\caption[]{The temperature dependences of the energy gaps $\Delta_1$ and $\Delta_2$. The full lines are the theoretical $\Delta(T)$ relations. Inset, the magnetic field dependence of $\Delta$ for $T=1.5~K$.}
\label{Fig2}
\end{figure}

We have measured IVC and its derivatives for heterocontacts between high-purity copper and slices ($1.5\times2\times3~mm^3$) of the ceramic $La_{1.8}Sr_{02}CuO_4$. The temperature of the start of the transition to the superconducting state of freshly prepared specimens was $T_{c1}=38~K$ and the width of the transition was 8~$K$. The scheme for carrying out the experiments is shown in the inset to Fig. \ref{Fig1}.
The choice of the tangent point at helium temperatures was made from the IVC corresponding to the S-c-N contact (Fig. \ref{Fig1}, curve 1) with a metallic behavior of the conductivity as a function of the applied bias ($d^2V/dI^2>0$), with excess current and with features on the derivatives \emph{reproducible for different contacts} which could be identified with the gap and the phonon energies. The energy gap for $T\ll T_c^*$ was calculated from the position of the minima in the first derivative (Fig. \ref{Fig1}, curves 2 and 3). The form of $dV/dI(V)$ transforms into that traditional for a S-c-N contact on raising the temperature (curve 4) and evaluation of the gap was made from the minima in the second derivative $d^2V/dI^2(V)$ (curve 5).

The temperature dependences of the values of the gap obtained in this way for of the contacts are shown in Fig. \ref{Fig2} (the points). Nonlinearity of the IVC first shows up for the temperature $T_c$ at $V=0$, produced by the transition of the region of the ceramic close to the contact into the superconducting state. For different contacts $T_c^*$ varied from zero to $\sim 32~K$, which indicates the nonuniformity in the distribution of the critical parameters on the surface of the specimen. Corresponding variations were also observed for the gap. The clearest difference between the observed variations from the theory (full curves) is the strong temperature dependence of both gaps in the region of temperatures low compared with $T_c^*$. It is possible that $\Delta_1$ and $\Delta_2$ correspond to different parts of the Fermi surface. If they go to zero at temperatures $T\Delta_1$ and $T\Delta_2$, then the superconductivity has a gapless character over a significant temperature range. The ratios $2\Delta_i(0)/k_BT_c^*$ are anomalously high (5.8 and 11.2 for $i=1$ and 2, respectively). On the one hand, this indicates an anomalously strong EPI, while on the other the inapplicability of the BCS theory in this case. In fact, the BCS theory assumes that the electron-phonon interaction (EPI) does not change at the transition to the superconducting state. Even in tightly bound superconductors, the value of the gap was
less than the characteristic frequencies of the low-energy phonons primarily responsible for the superconductivity. The energy gap in $La_{1.8}Sr_{0.2}CuO_{4}$ and in other similar compounds is so great that it is comparable with or even greater than the energy of low-frequency phonons. Moreover, the transition to the superconducting state evidently leads to a radical rearrangement of the electron spectrum, the EPI and the low-frequency phonon modes, which takes place over a wide range of temperature below $T_c^*$.

The diameter of our contacts, from various estimates, has a value from several tens to hundreds of \AA ngrstroms. For such small dimensions it is unlikely that the superconducting properties near the contact would be appreciably nonuniform and, consequently, the gaps $\Delta_1$ and $\Delta_2$ pertain to one and the same volume of the superconductor. From the ratio of the resistances of the contacts in the normal and
superconducting states, it can be deduced that the electron concentration in the ceramic is about 40 times less than in copper. The noticeable asymmetry of the IVC (Fig. \ref{Fig1}, curve 1) also indicates the smallness of the carrier concentration at the superconducting electrode.

The inset to Fig. \ref{Fig2} also shows the magnetic field dependence of the gap. The error in reading off the gap is small here, since the form of the $dV/dI$ characteristic in a magnetic field at low temperatures is practically constant and corresponds to curve 2 of Fig. \ref{Fig1}.
\begin{figure}[]
\includegraphics[width=8cm,angle=0]{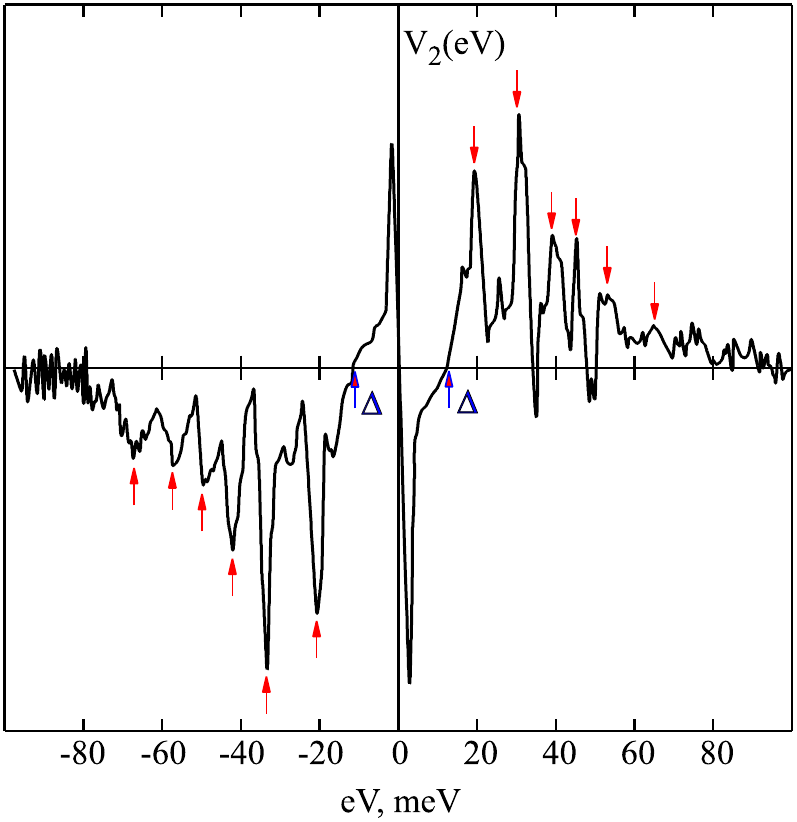}
\caption[]{Point contact spectrum of phonons in $La_{1.8}Sr_{0.2}CuO_{4}$.
The arrows indicate the positions of the maxima of the EPI.}
\label{Fig3}
\end{figure}

The point contact phonon spectrum is shown in Fig. \ref{Fig3}.
It extends to $\sim 70~meV$ and contains six lines (19, 32, 40, 47, 54, and 65~$meV$), corresponding to phons with small group velocities. The density of phonon states and the EPI function have maxima at these energies. The strength of  the EPI can be estimated from the relative growth in the differential resistance in the interval $\Delta_2<eV\leq \hbar\omega_{max}$ (the maximum phonon energy). The growth in $dV/dI$ because of inelastic scattering of electrons by phonons in the region of the contact reaches 50-100\% in the superconducting ceramic, while it is of the order of a few percent for typical metals. The metallic behavior of the conductivity over the whole energy range $eV\leq \hbar\omega_{max}$ is an indication of the smallness of thermal effects. The dependence of $dV/dI$ on $V$ above  $T_c^*$ has a semiconductor
character ($d^2V/dI^2<0$) in agreement with the negative temperature coefficient of the resistance at temperatures above the superconducting transition temperature $T_{c1}$. If a thermal regime were realized in the contact, then for $eV\cong 12~meV$ the temperature at the center of the contact would reach $T_{c1}$, and for large shifts of the IVC $d^2V/dI^2<0$ should obtain. In fact, however, a change in the sign of $d^2V/dI^2$ is observed on the IVC only for temperatures in the immediate vicinity of $T_c^*$. The transition to the superconducting state thus hinders the dielectrification of the spectrum and on further reduction of the temperature evidently assists the "metallization" of the electrons over a large part of the Fermi surface.
\\
\\
\textbf{NOTATION}
\\
\\
Here $T_{c1}$ is the superconducting transition temperature for the onset of bulk resistance, $T_c^*$ is the superconducting transition temperature for a point contact, $T\Delta_1$ and $T\Delta_2$ are the zero energy gaps extrapolated temperatures, $\hbar\omega_{max}$ is the maximum phonon energy, and $R_N$ is the point contact resistance in the normal state.

\end{document}